\documentclass[aps,prl,twocolumn]{revtex4}
\usepackage{epsfig,psfig}
\begin{document}

\title{Doping and Field-Induced Insulator-Metal Transitions in Half-Doped Manganites}
\author{O. C\'epas,$^{a,b}$ H. R. Krishnamurthy,$^{a,c}$ and T. V. Ramakrishnan$^{a,c,d}$}
\affiliation{
a. Centre for Condensed Matter Theory, Department of Physics, Indian Institute of Science, Bangalore 560012, India. \\
b. Institut Laue Langevin, BP 156, 38042 Grenoble, France. \\
c. Jawaharlal Nehru Centre for Advanced Scientific Research, Jakkur, Bangalore 560 064, India. \\
d. Department of Physics, Banaras Hindu University, Varanasi
221005, India.}

\date{\today}

\begin{abstract}
We argue that many properties of the half-doped manganites may be
understood in terms of a new two-($e_g$ electron)-fluid
description, which is energetically favorable at intermediate
Jahn-Teller (JT) coupling. This emerges from a competition between
canting of the core spins of Mn promoting mobile carriers and
polaronic trapping of carriers by JT defects, in the presence of
CE, orbital and charge order. We show that this explains several
features of the doping and magnetic field induced insulator-metal
transitions, as the particle-hole asymmetry and the smallness of
the transition fields.
\end{abstract}

\maketitle

\begin{figure}[htbp]
\vspace{-0.36cm}
\centerline{ \psfig{file=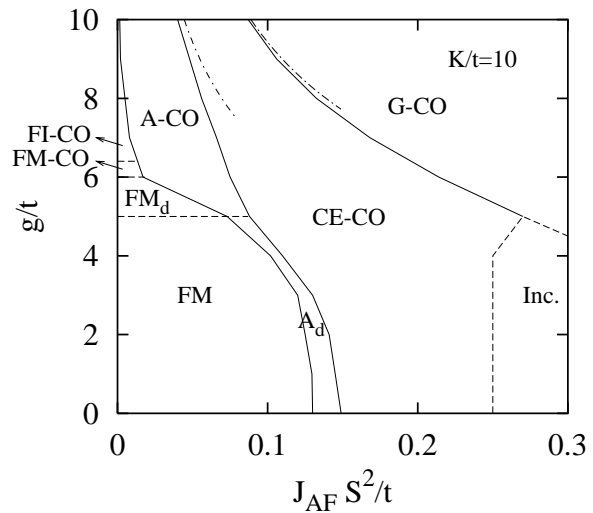,width=7cm,angle=-90}}
\vspace{-0.1cm}
\caption{Phase diagram of the 3D two-orbital model ($T=0$,
$x=0.5$, $K/t=10$). FM (resp. FM$_d$): ferromagnetic metallic
phase with no distortions (resp. small uniform distortions). FI-CO
(resp. FM-CO): charge-ordered ferromagnetic insulating (resp.
metallic) phase with distortions that favor occupancy of the
$x^2-y^2$ orbitals. A$_d$: ferromagnetic planes AF aligned with
uniform distortions. A-CO: A with charge order.  CE-CO:
Ferromagnetic zig-zag chains AF ordered, orbital ordered
($3x^2-r^2/3y^2-r^2$ on the bridge sites), and charge-ordered
($g/t>0$). G-CO: N\'eel AF phase with charge-order. Inc.:
incommensurate state that interpolates between CE and G. Dotted
dashed lines come from analytical expressions derived in the
strong-coupling limit. Solid (dashed) lines show first-order
(second-order) phase transitions.} \label{pd-f}
\end{figure}

``Half-doped'' manganites such as Re$_{1-x}$A$_{x}$MnO$_3$ with
$x=1/2$ where Re is a 3+ rare-earth ion and A a 2+ alkaline earth
ion have been the object of extensive studies for many years
\cite{Salamon}. The lowest temperature phase seems to be either
the CE phase, consisting of ferromagnetic {\it zig-zag chains}
with relative antiferromagnetic (AF) order (as in
La$_{1/2}$Ca$_{1/2}$MnO$_3$, where it was first proposed
\cite{Wollan,Goodenough}, and in Nd$_{1/2}$Sr$_{1/2}$MnO$_{3}$
\cite{Kawano} or Nd$_{1/2}$Ca$_{1/2}$MnO$_3$ \cite{Millange}) or
an A-type phase, i.e, ferromagnetic {\it planes} with relative AF
alignment  (as in Pr$_{1/2}$Sr$_{1/2}$MnO$_3$ \cite{Kawano}).  The
competition between the CE and A phases appears even in a simple
one-orbital model \cite{Pandit} because of the interplay of
ferromagnetic double-exchange and AF superexchange between the
core $t_{2g}$ spins of Mn (see also Fig.  \ref{pd-f}). The
presence of charge and orbital order as proposed by Goodenough
\cite{Goodenough} is more difficult to establish. X-ray
diffraction experiments do suggest the presence of large
Jahn-Teller (JT) distortions \cite{Radaelli,Kawano} with two
inequivalent Mn sites. In the CE phase, the alternating
$(3x^2-r^2)/(3y^2-r^2)$ orbital order (consistent with the
observed distortions) was shown to optimize the anisotropic
hopping energy of the $e_g$ electrons in a more realistic two
$e_g$ orbital model \cite{Khomskii}. The origin of charge-order
was attributed to on-site \cite{Khomskii} or intersite Coulomb
interactions \cite{Jackeli,Fratini}, though the latter tends to
favor a Wigner crystal \cite{Jackeli} rather than the charge
stacked order found experimentally \cite{Salamon}.  The role of
the JT coupling has been investigated using imposed JT distortions
\cite{Popovic} as well as by extensive classical Monte-Carlo
simulations that lead to the observed charge stacked
ordering \cite{Dagotto}.

However, several fundamental issues remain to be understood. One of
them is the striking asymmetry with respect to the addition of
electrons or holes. Experimentally, added electrons typically favor
ferromagnetic metallic phases while added holes favor insulating
phases \cite{Salamon}. In contrast, band structure arguments
\cite{Khomskii}, and treatments including JT distortions adiabatically
and classically \cite{Dagotto} lead to metallic phases on both sides.
Another puzzling feature, first seen in
(Nd,Sm)$_{1/2}$Sr$_{1/2}$MnO$_3$ \cite{Tomioka} and later seen to be
ubiquitous \cite{Salamon}, is that magnetic-fields $\sim 10-40$ Tesla,
which are extremely small compared with N\'eel or charge ordering
temperatures $\sim 200$ K, induce an insulator-metal transition. This
can be viewed as another manifestation of the colossal
magneto-resistance (CMR) in doped manganites \cite{Salamon}.
An explanation is that this arises from the proximity of the CE
phase to a ferromagnetic phase \cite{Pandit,Fratini,Aliaga};
but it is difficult to understand why the parameters in so many
systems should all be so finely tuned as to be near the phase
boundary.

Recently, starting from a large JT coupling picture, a two-fluid
$e_g$ electron model, one polaronic and localized, and the other
band-like and mobile, was proposed and shown to explain, in
particular, the CMR in the orbital liquid regime \cite{Ram}. In
this letter, we show how the two types of electrons can emerge
from a realistic microscopic model, even at intermediate JT
couplings, in the half-doped case where orbital and charge order
have to be explicitly included.  Basically, they arise from a
competition between canting of the Mn core spins promoting mobile
carriers, and the JT coupling promoting polaronic, localized
carriers. We show that our picture leads to natural explanations
for the particle-hole asymmetry around half-doping as well as the
magnetic-field-induced insulator-metal transition at half-doping
mentioned above. Interestingly, a similar two-carrier-type
hypothesis was proposed {\it based on phenomenological grounds} in
Ref. \onlinecite{Roy} to understand resistivity data in
La$_{1-x}$Ca$_{x}$MnO$_3$ ($x \sim 1/2$); for which our theory
provides a microscopic basis. JT distortions were recently tracked
as function of field in La$_{1/2}$Ca$_{1/2}$MnO$_3$ and shown to
play a crucial role near the field-induced transition
\cite{Tyson,Nojiriprivate}; our picture is completely consistent
with this. We believe that the ideas presented here may be
relevant to other classes of systems such as CsC$_{60}$, in which
a similar two-electron phase has been proposed \cite{Alloul}.

Our theory is based on the following microscopic two-orbital
Hamiltonian for the manganites:

\begin{eqnarray}
{\cal H}[\{ \textbf{\mbox{S}}_{i a} , Q_{i a},\Theta_{i a} \} ] =
- \sum_{ij \alpha \beta ab} \tilde{t}_{ab ij}^{\alpha \beta}
(\textbf{\mbox{S}}_{i a},\textbf{\mbox{S}}_{j b}) c^{\dagger}_{i a
\alpha } c_{j b \beta} \nonumber \\+ \sum_{<ia,jb>} J_{AF}
\textbf{\mbox{S}}_{i a} \cdot \textbf{\mbox{S}}_{j b} - g \mu_B
\sum_{i a} \textbf{\mbox{H}}
\cdot \textbf{\mbox{S}}_{i a}  \nonumber \\
+ \frac{1}{2} K \sum_{i a} Q_{i a}^2 - g \sum_{i a \alpha \beta }
Q_{i a}   c_{i a \alpha }^{\dagger} \tau_{\alpha \beta}(\Theta_{i
a}) c_{i a \beta}.\label{Hamiltonian}
\end{eqnarray}
Here $c^{\dagger}_{i a \alpha }$ creates an electron in the $e_g$
orbital $\alpha$ ($=x^2-y^2$, $3z^2-r^2$) in the unit cell $i$ and
a sublattice site labelled by $a$. (We use a 8-sublattice
decomposition to accommodate the CE phase.) There are $N$ sites
and $cN$ electrons with $c \equiv (1-x)$ close to $1/2$. Due to a
large Hund's coupling $J_H$ the electron spin is assumed to be
locked parallel to the $S=3/2$ $t_{2g}$ core spins of Mn, modelled
as {\it classical vectors} $\textbf{\mbox{S}}_{ia}$. The hopping
parameters (with $4t/3$ being the hopping between $(3z^2-r^2)$
orbitals in the $z$-direction) include the standard
Anderson-Hasegawa dependence on $\textbf{\mbox{S}}_{ia},
\textbf{\mbox{S}}_{jb}$  that takes care of this large $J_H$
projection \cite{Dagotto}. The core spins are directly coupled by
an AF superexchange, $J_{AF}S^2 \sim 0.1 t $ \cite{Dagotto}.
$\textbf{\mbox{H}}$ is the external magnetic field. The last two
terms include the vibrational energy of JT phonons (where $K$ is
the lattice stiffness of a simplified non-cooperative model) and
their coupling to the $e_g$ electrons. $Q_{i a }$ and $\Theta_{i
a}$ represent the amplitude and the angle of the two ($Q_2,Q_3$)
JT modes, and the $\tau$ matrix the symmetry of their coupling
\cite{Dagotto}.  On-site Coulomb interactions can be ignored in a
first approximation when large JT distortions are present (as the
JT coupling suppresses double occupancy) and for large $J_H$.

We have determined the ground state of (\ref{Hamiltonian}) exactly
numerically, {\it but in the subspace of spin and distortion
variables restricted to be periodic with a unit cell of at most 8
sites}. This accommodates the CE state as well as several other
competing commensurate states. Compared to earlier numerical
approaches \cite{Dagotto} that were limited to small clusters, our
calculations are practically in the thermodynamic limit. We
confirm the phase diagram that was previously obtained
\cite{Dagotto,Aliaga} and obtain detailed predictions on the
strength of the JT-distortions $Q$, etc. \cite{long-paper}. The
phase diagram is given in Fig. \ref{pd-f} and the phases are
described in the figure caption.
The strong-coupling phases, all insulating and charge ordered, can
be understood by starting from localized Wannier orbitals centered
on alternate JT distorted sites which are fully occupied. By {\it
virtual double exchange} involving neighboring empty sites with
aligned core spins \cite{Ram}, the electrons gain energy in a way
that depends upon the orientation of the JT distortion or occupied
orbital \cite{long-paper}. A comparison of the energies of the
various phases leads to the sequence of first-order transitions at
couplings given by $J_{AF}S^2=4tK/(9g^2)$ and
$J_{AF}S^2=8tK/(9g^2)$ (dotted dashed lines in Fig. \ref{pd-f}).

Phases that are inhomogeneous or incommensurate \cite{incom-CO}
can not be captured by the above analysis because of the limited
size of the maximal unit-cell. We tackle this problem, albeit to a
limited extent, by studying the {\it instabilities} of the
homogeneous insulating phases discussed above with respect to
particle or hole excitations {\it accompanied by single site
defects in their JT distortion pattern}. For this, we find the
electronic eigenvalues of (\ref{Hamiltonian}) in the {\it presence
of such defects} numerically (with $N$ up to 1728), and calculate
the energy cost or gain from filling the energy levels with $cN$
electrons.

\begin{figure}[tbp]
\vspace{-0.36cm}
\centerline{\psfig{file=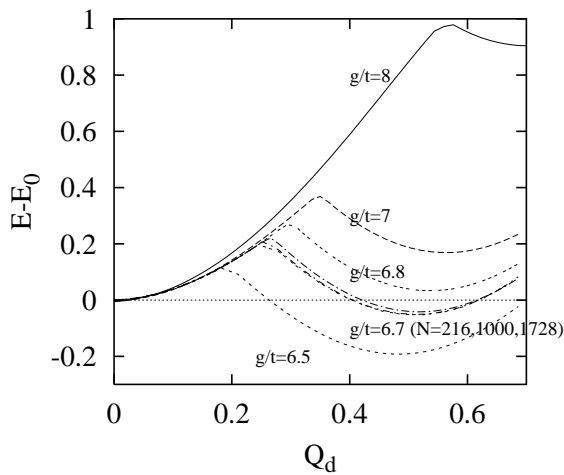,width=6.5cm,angle=-90}}
\vspace{-0.1cm}
\caption{Energy change when a single JT defect is introduced in the
FI-CO phase. $Q-Q_d$ is the JT distortion on a defect site; all the
other occupied sites having the same distortion $Q$. The softening of
the excitation with $Q_d \sim Q$ at $g/t \sim 6.8$ signals a phase
transition with proliferation of defects.  Finite-size effects are
small and shown for $g/t=6.7$.} \label{deltaelastic-f}
\end{figure}

To start with, consider the FI-CO phase at strong-coupling, with
the electrons localized at the JT distorted sites with distortion
$Q$. If we now promote a particle across the charge gap, it is
energetically favorable for the JT distortion at the hole site
(from which the electron is removed) to relax to $Q-Q_d$. The loss
in electronic energy due to the scattering of the other electrons
from the defect is overcompensated by the gain in elastic energy.
The hole gets polaronically trapped, while the electron is mobile.
Such mixed excitations thus have energies lower than the energies
of particle-hole excitations due to band structure alone. For the
FI-CO phase, this is demonstrated in Fig. \ref{deltaelastic-f}
where, in addition to the minimum at $Q_d=0$ (corresponding to the
homogeneous phase), there is another minimum at $Q_d \sim Q$,
corresponding to the removal of the JT distortion at one site.
Furthermore, this minimum softens when $g/t$ is reduced below $g/t
\sim 6.8$ (Fig. \ref{deltaelastic-f}), although the other
homogeneous phases of Fig. \ref{pd-f} are at higher energy at this
$g/t$. The instability approach therefore suggests that there
might be another phase where such defects are energetically
favorable and proliferate.
With a small number of {\it the above type of defects}, a small
fraction of electrons are converted from localized to mobile
states leading to a metal with a small concentration of mobile
electrons. This is reminiscent of the two-fluid picture
\cite{Ram}, but now extended to accommodate orbital and charge
order \cite{long-paper}.

\begin{figure}[tbp]
\vspace{-0.36cm}
\centerline{ \psfig{file=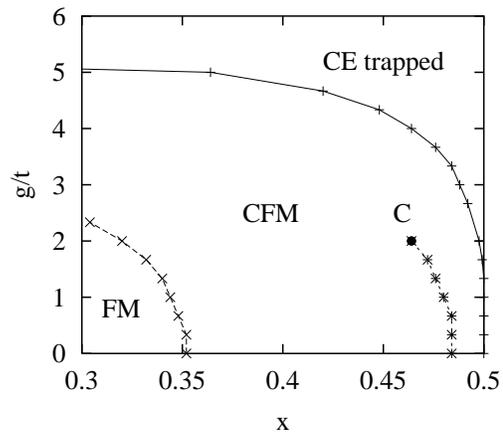,width=6cm,angle=-90}}
\vspace{-0.1cm}
\caption{Phase diagram, $g/t$ vs. doping, $x$
($J_{AF}S^2/t=0.15$). CFM: canted CE state with distortions and
charge-order (metallic). CE trapped: CE phase with extra carriers
trapped in JT distortions (insulating).  FM: ferromagnetic
metallic phase with no distortions. The upper curve is valid for $x$
close enough to $0.5$. C is a critical point
ending a first-order line between two canted states (with
different canting angles).}
\label{Phase diagram-f}
\end{figure}

We next address similar instability issues in the context of the CE
phase. First consider what happens when the CE phase is doped with
carriers. As discussed above, experimentally there is a strong
asymmetry between hole and electron doping. According to de Gennes's
original argument \cite{deGennes}, canted phases are expected for
small doping (irrespective of their sign). It is known that the energy
of the fully ferromagnetic state crosses that of the CE state when
extra electrons are added \cite{Khomskii}, but intermediate canted
phases have not been considered. They would naively lead to second-order
transitions rather than first-order.  We have studied such canted
phases and obtained the optimal canting angle
%compared the energies of the CE,
%CE-canted and fully FM phases 
as function of $x$ close to 1/2. Similarly to the discussion above in
the context of the FI-CO phase, these homogeneous phases are in
competition with inhomogeneous phases where the added carriers are
self-trapped by JT distortions. In fact, we find that it is favorable
to trap the added \textit{carriers} at small $\delta c \equiv 1/2-x$.
This leads to an insulating un-canted CE phase (noted CE trapped in
Fig. \ref{Phase diagram-f}). On the \textit{electron}-doped side
($x<1/2$, Fig. \ref{Phase diagram-f}), increasing $\delta c$ leads
eventually to a canted metallic phase (CFM) via a first-order
transition.  This is because the JT energy gain due to trapping is linear in
$\delta c$, $\delta E_{tr} = - \tilde{E}_{JT}^{e,h} |\delta c|$ (with
$\tilde{E}_{JT}^{e,h}$ obtained by solving, for all $g/t$, the
one-defect problem mentioned above with one added carrier)
\cite{long-paper}, whereas the energy gain from canting is quadratic,
$\delta E_{ca} \sim -(\delta c)^2$ \cite{deGennes}. The latter loses
for small $\delta c$ but wins for larger $\delta c$.
On the \textit{hole} side, however, we find that canted phases are
never energetically favorable. The asymmetry arises because of the
nature of the CE ordering. Canting leads to a 2-d dispersion, with a
large density of states at the bottom of the conduction band (for
electron doping), whereas it gives a 3-d dispersion, with a vanishing
DOS at the top of the valence band (for hole doping). Therefore,
canting angles can indeed get large when electrons are added and
compete effectively against electron trapping. But when holes are
added, canting angles are much smaller and the holes get trapped
by JT distortions for $g/t>4$.  Hence the system remains
insulating. Thus, our approach leads to an explanation for the
asymmetry between particle and hole doping seen experimentally. It
also helps us to understand why incommensurate charge ordered CE type
phases seem to be favoured on the hole doped side
\cite{incom-CO,long-paper}.

\begin{figure}[htbp]
\vspace{-0.36cm}
\psfig{file=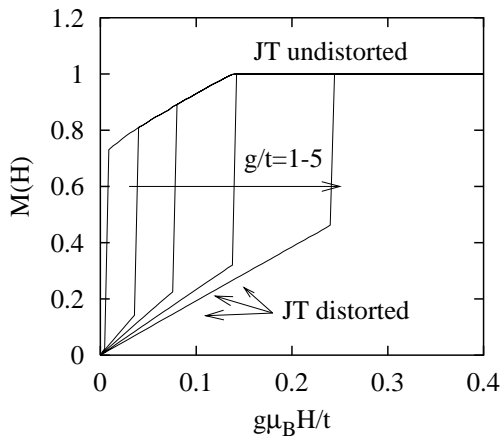,width=6cm,angle=-90}
\vspace{-0.1cm}
\caption{Magnetization vs. field ($g/t=1-5$,
 $J_{AF}S^2/t=0.15$). The first-order transitions to an
\textit{undistorted} highly canted (or fully FM) metallic
phase is accompanied with a relaxation of the JT distortions.
} \label{magnetizationgap-f}
\end{figure}

An external magnetic-field applied to the CE phase also promotes
canting. Experimentally, as discussed earlier, a field-induced
insulator-metal transition occurs at extremely small fields. To locate
the transition in our theory, we minimize and compare the energies of
various 8-sublattice structures \textit{in a field}, including the
JT-\textit{distorted} canted CE state, the \textit{undistorted} canted
state with the optimal (high) canting angle, etc.  We find that for $
g/t < 5$ the ground state switches in a first-order transition from a
distorted canted CE phase (with the canting hardly changing the JT
distortions) to an undistorted highly-canted (or FM) phase with
increasing field. At the transition, the system becomes metallic,
there is a jump in the magnetization (Fig.  \ref{magnetizationgap-f}),
and an abrupt relaxation of all the JT distortions to zero. The
transition fields have very little to do with the magnetic energy
scales, but are determined by the JT energies and depend strongly on
$g/t$ as is clear from Fig.  \ref{magnetizationgap-f}. For $g/t
\gtrsim 6.8$, the ferromagnetic state is insulating and no
insulator-metal transition can be found, which puts a bound on the
values of $g/t$ that are appropriate. In the range $ 5.0 \lesssim g/t
\lesssim 6.8$, which may be relevant for manganites (we need $g/t
\gtrsim 5.0$ to explain the existence of the A-CO phase \cite{Kawano}
[see Fig. \ref{pd-f}]), we find an instability of the distorted canted
CE phase towards creation of defects, which suggests that the
field-induced metallic phase in this case has the above mentioned two
types of electrons. In all cases, there are abrupt changes in JT distortions 
at the transition, converting it from second-order (for progressive
canting) to first-order, in agreement with recent experiments
\cite{Tyson,Nojiriprivate}.

However, the transition fields obtained in our calculations are
too large compared to experiments. For instance, $g \mu_B H_c \sim
0.1 t$ (Fig. \ref{magnetizationgap-f}), gives $H_c \sim 140$ T
(with $t \sim 0.2 $ eV). The discrepancy is connected with the
overestimation of the charge gap in our model. Three effects need
to be included to obtain a more realistic, reduced estimate for
the charge gap. First, the finiteness of the Hund's coupling, here
taken to be infinite, which would allow for hopping even between
sites with anti-aligned core spins; second, the cooperative nature
of the JT phonons, causing sizeable distortions on the corner
sites of the CE phase as well; and third, small second neighbour
hopping. All of these would contribute to reducing the gap.
Indeed, if we use the experimentally observed charge gap in place
of the charge gap obtained in our model and then estimate the
transition field, we get numbers in good agreement with
observations.

In conclusion, we have provided new theoretical insights into the
physics of half-doped manganites. We suggest the existence of and
competition between canting (i.e., not full ferromagnetism, which
could be checked by neutron diffraction) induced metallicity and
inhomogeneity arising from the trapping of carriers
by JT defects. It explains several features of the doping-induced
(e.g. the particle-hole asymmetry) and field-induced insulator-metal
transition.
%At the intermediate JT
%couplings which we believe is appropriate for manganites, the
%metal has JT defects. The majority of electrons are then
%polaronically trapped and localized. A small fraction of the
%electrons are promoted to the conduction band and are mobile,
%leading to a two-fluid picture. This is consistent with the
%findings in La$_{1/2+x}$Ca$_{1/2-x}$MnO$_3$ in ref.
%\onlinecite{Roy}. The IM transition is accompanied by large
%readjustments of the JT distortions, which nevertheless survive at
%a large fraction of sites. This is consistent with X-ray
%measurements under field \cite{Tyson,Nojiriprivate}.
These ideas suggest a new two-fluid model with localized and mobile
electrons, which extends the work of ref. \onlinecite{Ram} to include
orbital and charge order, which when treated with more sophisticated
methods such as DMFT can yield a satisfactory and complete theory of
doped manganites including the regime near half-doping.

O.C. would like to thank G. Bouzerar, T. Chatterji, G. Jackeli, D.
Khomskii, Y. Motome, H. Nojiri and T. Ziman for stimulating
discussions. OC and HRK acknowledge financial support from IFCPAR, grant
2404-1.

\end{document}